\documentclass[11pt,a4paper]{article}
\usepackage{graphicx}
\usepackage{mathtools}
\usepackage{amsfonts}
\usepackage{amssymb}
\usepackage{amsmath}
\usepackage{bm}
\usepackage{hyperref}
\usepackage{titlesec}

\begin{document}
	\title{The unification of gravity and the spin-1 field}
	\author{Gary Nash
		\\University of Alberta, Edmonton, Alberta, Canada,T6G 2R3\\gnash@ualberta.net \footnote{Present address Edmonton, AB.}
	}
	\date{~Dec 20, 2024}
	\maketitle
	\vspace{2mm}

\begin{abstract}
Unifying the massive spin-1 field with gravity requires the implementation of a regular vector field that satisfies the spin-1 Proca equation and is a fundamental part of the spacetime metric. That vector field is one of the pair of vectors in the line element field (\textbf{X},-\textbf{X}), which is paramount to the existence of all Lorentzian metrics and Modified General Relativity (MGR). Symmetrization of the spin-1 Klein-Gordon equation in a curved Lorentzian spacetime introduces the Lie derivative of the metric along the flow of one of the regular vectors in the line element field. The Proca equation in curved spacetime can then be described geometrically in terms of the line element vector, the Lie derivative of the Lorentzian metric, and the Ricci tensor, which unifies gravity and the spin-1 field. Related issues concerning charge conservation and the Lorenz constraint, singularities in a spherically symmetric curved spacetime, and geometrical implications of MGR to quantum theory are discussed. A geometrical unification of gravity with quantum field theory is presented.\vspace{3mm}

\end{abstract}
\vspace*{0.4cm}
\newline This is a preprint of an article published in International Journal of Modern Physics A. The original authenticated version is available online at: https://doi.org/10.1142/S0217751X25500058. 

\section{Introduction}
After Einstein \cite{Ein16} published the final form of General Relativity (GR) in 1916, he spent most of the last 30 years of his life trying to unify gravity with electromagnetism. Several, but not all, of his approaches to that unification are now briefly discussed. In his 1925 paper \cite{Ein25} on the subject, he introduced a metric and a connection, which were both asymmetric, to obtain an antisymmetric tensor that represents the electromagnetic field within the geometrical description of gravity by GR. However, the recovered electromagnetic field equations in a weak gravitational field were not entirely equivalent to Maxwell's equations. Not having achieved the desired success with that approach, he introduced an asymmetric connection where torsion was implicit in his 1930 paper \cite{Ein30} on teleparallelism but did not succeed in deriving equations of motion for particles. He then concentrated on the antisymmetric part of the metric in his 1945 and 1946 papers \cite{Ein45,Ein46} but felt that this approach to unification yielded a theory that was considered unified only in a limited sense. He and Straus did not obtain solutions to the linearized equations in the 1946 paper that were rigorous and regular in the entire space.
\par Another approach to the unification of gravity and electromagnetism was presented in 1957 by Misner and Wheeler \cite{Wheel} based on the \emph{conditions} that $R=0$ and $R^{\beta}_{\alpha}R^{\gamma}_{\beta}=\frac{\delta^{\gamma}_{\alpha}}{4}R_{\sigma\rho}R^{\sigma\rho}$. Einstein's and Maxwell's equations, each with second-order derivatives, were replaced by one set of equations involving fourth-order derivatives that were purely geometrical. This is known as the ``already unified field theory" or geometrodynamics. More recently it was shown in Ref. \cite{Lind} that by \emph{assuming} the metric could be expressed as the product of the electromagnetic four-potentials $A_{\alpha}$, it is possible to express the Ricci scalar in terms of $A_{\alpha}$ and its derivatives up to second order. Spacetime was required to be Ricci-flat to be consistent with Maxwell's equations.
\par  These particular references pertain to the attempts to unify gravity with Maxwell's equations of electromagnetism in a four-dimensional spacetime. More generally, it was Einstein's dream to construct a geometrical theory of gravity that is unified with quantum theory in a singularity-free continuous curved spacetime. The unification of GR and quantum theory is paramount to establishing the \emph{basis} for a theory of quantum gravity: the quantization of gravity or the geometrization of quantum theory; unification in the sense that each theory can be described geometrically and contains the same geometrical entity, which links the theories. This entity is the Lie derivative of a Lorentzian metric along a regular (everywhere nonvanishing) vector field that satisfies the spin-1 Proca equation in curved spacetime and is a fundamental part of the Lorentzian metric. The regular vector field that explicitly belongs to that metric is the line element field, which property of Lorentzian metrics is scarcely known but has been used in GR to develop theorems \cite{CB,HE} on causality. Thus, this paper focuses on the geometrical implications of introducing the Lie derivative of a particular Lorentzian metric along the line element vector that belongs to that metric, into quantum theory. This leads to the geometrical unification of gravity and the spin-1 field, and geometrical explanations of particular mysteries of quantum theory. The line element vector field is now reviewed.
\section*{The line element vector field}
As stated in Ref. \cite{Markus}, every paracompact differentiable manifold $ \mathcal{M} $ admits a continuous non-zero vector field and thus a continuous line element field without singularities. That means there exists a differentiable regular vector field $ \bm{X} $ collinear to a differentiable non-vanishing unit vector $ \bm{u} $ by $\bm{X}=f\bm{u}  $ where $ f\neq0 $ is the scalar magnitude of $ \bm{X}$. The line element (or direction) field $(\bm{X},-\bm{X}) $ is defined as an assignment of a pair of equal and opposite vectors at each point of $\mathcal{M}$. It is a one-dimensional vector subspace of the tangent space on $ \mathcal{M} $. \par The line element field is fundamental to the existence of a Lorentzian metric. As proved in Refs. \cite{CB,HE,Markus}, a paracompact manifold admits a Lorentzian metric $ g_{\alpha\beta} $ if and only if it admits a line element field. A Lorentzian metric with a +2 signature is constructed from a Riemannian metric $ g^{+}_{\alpha\beta} $, which always exists on $\mathcal{M}$, and a product of unit line element covectors associated with  $ \bm{u} $, as 
\begin{equation}\label{gab}
	g_{\alpha\beta}=g_{\alpha\beta}^{+}-2u_{\alpha}u_{\beta}
\end{equation}where the Riemannian metric $g^{+}_{\alpha\beta}$ is independent of the covectors $u_{\alpha}$ and the unit vectors satisfy $g_{\alpha\beta}u^{\alpha}u^{\beta}=-1$, and $g^{+}_{\alpha\beta}u^{\alpha}u^{\beta}=1$. Thus, a Lorentzian metric does not exist without the line element field. The regular vectors in the line element field are not introduced as absolute vectors in a Lorentzian spacetime $(\mathcal{M},g_{\alpha\beta})$ because it does not exist without them. \par As noted in Ref. \cite{ND}, a Lorentzian metric is highly nonunique; it contains an arbitrary Riemannian metric and an arbitrary pair of line element covectors. Nevertheless, a particular Lorentzian metric exposes a unique pair of covectors dual to the collinear vectors \textbf{u} and
\textbf{X} by g. This is how the word ``particular" is to be interpreted when it refers to a Lorentzian metric in this article. The arbitrary Riemannian metric $g^{+}$ in a particular Lorentzian metric $g$ gives it the flexibility to describe a myriad of geometries. \par The physical interpretation of the line element field is now discussed. The line element vectors at each point of a Lorentzian spacetime can be divided into timelike, spacelike, and null classes depending on whether $X_{\beta}X^{\beta}$ is negative, positive or zero, respectively. The null vectors form the familiar double cones in the tangent space on $\mathcal{M}$, which separates the timelike vectors from the spacelike vectors. A Lorentzian metric has a directional characteristic that is described by the line element vector associated with the metric.\par The line element vector field is a fundamental part of the elusive connection-independent symmetric tensor $\varPhi_{\alpha\beta}$ constructed from the Lie derivative of a particular Lorentzian metric along the flow of the line element vector, which belongs to that metric. That tensor completes the Einstein equation and describes the local energy-momentum of the gravitational field. The geometric development of $\varPhi_{\alpha\beta}$ is now shown from a review of Modified General Relativity (MGR) discussed in Refs. \cite{ND,N}.
\section*{Modified General Relativity}
Einstein developed his equation of General Relativity 
\begin{equation}\label{EIN}
	G_{\alpha\beta}=\frac{8\pi G}{c^{4}}T_{\alpha\beta}
\end{equation}in a four-dimensional Riemannian spacetime with the understanding that spacetime is locally Minkowskian during free-fall. He realized that his equation should contain a symmetric tensor that describes the energy-momentum of the gravitational field but was unable to construct that tensor. At the time of his death in 1955, knowledge of Lorentzian metrics  \cite{Markus} was progressing in the study of pseudo-Riemannian geometry. In 1969, Berger and Ebin \cite{Berger} provided a theorem on the orthogonal decomposition of (0,2) tensors in the symmetric cotangent bundle $S^{2}T^{*}\mathcal{M}$ on compact Riemannian manifolds. That theorem was adapted to paracompact Riemannian manifolds without boundary and with smooth sections of $S^{2}T^{*}\mathcal{M}$ with compact support in Ref. \cite{Gil}, which was extended to paracompact Lorentzian manifolds in Refs. \cite{N,ND} to produce the Orthogonal Decomposition Theorem (ODT):\newline An arbitrary non-divergenceless (0,2) symmetric tensor $ w_{\alpha\beta} $ in the symmetric cotangent bundle $ S^{2}T^{\ast}\mathcal{M} $ with smooth sections of compact support on an n-dimensional paracompact boundaryless time-oriented Lorentzian manifold $ (\mathcal{M},g_{\alpha\beta}) $ with a Levi-Civita connection can be orthogonally decomposed as $
w_{\alpha\beta}= v_{\alpha\beta}+ \varPhi_{\alpha\beta} $ where $v_{\alpha\beta}  $ represents a linear sum of symmetric divergenceless (0,2) tensors and $\varPhi_{\alpha\beta}=\frac{1}{2}\pounds_{X}g_{\alpha\beta}+\pounds_{X}u_{\alpha}u_{\beta}$ where the timelike unit vector $\bm{u}  $ is collinear with one of the pair of regular vectors in the line element field $ (\bm{X},-\bm{X}) $ and $\bm X$ is not a Killing vector.\newline In a four-dimensional spacetime, the linear sum of symmetric divergenceless (0,2) tensors in the ODT contains the metric and the Einstein tensor as a consequence of Lovelock's theorem \cite{Love}:\newline In a four-dimensional spacetime, the only tensors which are symmetric, divergenceless, and a concomitant of the metric together with its first two derivatives, are the metric and the Einstein tensor.\par 
Einstein believed that gravity gravitates and postulated\cite{Ein16,EinGros} that the total energy-momentum tensor $T_{\alpha\beta}$ consisted of the energy-momentum from ponderable (ordinary) matter plus that of the gravitational field. This demands the local conservation of total energy-momentum with the vanishing of $ \nabla^{\alpha}T_{\alpha\beta} $ and not $ \nabla^{\alpha}\tilde{T}_{\alpha\beta} $, where $ \tilde{T}_{\alpha\beta} $ is the ordinary matter component of $ T_{\alpha\beta} $. Since $ \tilde{T}_{\alpha\beta} $ is not divergenceless, it can be set with the constant $k$ to an arbitrary non-divergenceless symmetric tensor $ w_{\alpha\beta} $, which is then orthogonally decomposed by the ODT: $w_{\alpha\beta}=k\tilde{T}_{\alpha\beta}=v_{\alpha\beta}+\varPhi_{\alpha\beta} $. Since $T_{\alpha\beta} $ is uniquely determined up to a divergenceless tensor, the sum of possible divergenceless non-Lovelock tensors in $v_{\alpha\beta}$ can be ignored. Hence, $v_{\alpha\beta}=\Lambda g_{\alpha\beta}+G_{\alpha\beta}$ where $\Lambda$ is the cosmological constant. The ODT and Lovelock's theorem generate the \emph{complete} Einstein equation \emph{in one line} 
\begin{equation}\label{MEQ}
	\Lambda g_{\alpha\beta}+G_{\alpha\beta}+\varPhi_{\alpha\beta}=k\tilde{T}_{\alpha\beta}
\end{equation}
with $k:=\frac{8\pi G}{c^{4}}$. This is the geometric basis of Modified General Relativity; it is the completion of General Relativity with the symmetric tensor describing local gravitational energy-momentum that Einstein originally proposed but could not construct.
\par It took over a century to produce the tensor that Einstein sought. Theoretical physicists generally considered constructing that tensor within the structure of GR to be an
impossible task \cite{PWill}. Some theorists dismissed the need for such a tensor because GR is nonlinear. By expanding the Ricci tensor to second order in a perturbed metric, an energy-momentum tensor that depends only on the background gravitational field and contains terms quadratic in the perturbed metric and its first two covariant derivatives can be constructed \cite{Isaa,Magg}. However, that tensor is not connection-independent, and local gravitational energy-momentum is prohibited by the equivalence principle.
\par MGR exploits a particular Lorentzian metric to construct $ \varPhi_{\alpha\beta} $ from the
ODT: 
\begin{equation}\label{Phiab}
	\begin{split}
		\varPhi_{\alpha\beta}:=	\frac{1}{2}\pounds_{X}g_{\alpha\beta}+\pounds_{X}(u_{\alpha}u_{\beta}).
	\end{split}
\end{equation} The covectors $u_{\alpha}$ and $u_{\beta}$ in (\ref{gab}) are precisely those in (\ref{Phiab}). The flow vector $\textbf{X}$ in the Lie derivatives of (\ref{Phiab}) is collinear with \textbf{u}. Thus, the vectors \textbf{u} and \textbf{X} employed in MGR are directly related to a particular Lorentzian metric and are not introduced arbitrarily. With $\textbf{u}$ in an orthonormal basis,
\begin{equation}\label{Phi00}
	\varPhi_{00}=(1+2u_{0}u^{0})\nabla_{0}X_{0}
\end{equation}and the local gravitational pressure is
\begin{equation}\label{Xi}
	\varPhi_{ii}=(1+2u_{i}u^{i})\nabla_{i}X_{i}\;\;\text{no sum on i},\;\;i=1,2,3.
\end{equation}$\varPhi_{\alpha\beta}  $ vanishes if and only if $ X^{\mu} $ is a Killing vector. However, in general, there are no Killing vector fields unless a particular symmetry is involved.
\par  $\varPhi_{\alpha\beta}$ is constructed with the Lie derivative and is therefore independent of the Levi-Civita connection $\nabla$. When the connection coefficients vanish under free fall, $ \varPhi_{\alpha\beta} $ is equivalently expressed with partial derivatives. That property is \emph{essential} to the existence of local energy-momentum relative to the equivalence principle because local gravitational energy-momentum does not vanish during free fall and mysteriously reappear after the event; it is invariant under free fall. Local gravitational energy-momentum is defined by $\varPhi_{\alpha\beta}$ and there is no conflict with the equivalence principle.
\par  Equation (\ref{MEQ}) can be obtained dynamically from the action functional $ S=S^{F}+S^{EH}+S^{G} $, which consists of the action for all ordinary matter fields $S^{F}$, the Einstein-Hilbert action of GR, $S^{EH}$, and the action for the energy-momentum of the gravitational field, $S^{G}$, that is defined in terms of the scalar $\Phi$, the trace of $\varPhi_{\alpha\beta}$ with respect to the inverse metric:
\begin{equation}\label{S}
	\begin{split}
		S=\int L^{F}( A^{\beta},\nabla^{\alpha} A^{\beta},...,g^{\alpha\beta})\sqrt{-g}d^{4}x
		\\+\frac{c^{3}}{16\pi G} \int (R-2\Lambda)\sqrt{-g}d^{4}x-\frac{c^{3}}{16\pi G}\int \varPhi_{\alpha\beta} g^{\alpha\beta}\sqrt{-g} d^{4}x
	\end{split}
\end{equation}
where $L^{F}  $ is the Lagrangian of the ordinary matter fields $A^{\beta}$. The details of the variations of $S$ with respect to the variables $g^{\alpha\beta}$, $ u^{\nu}$ and $f $ are given in Ref. \cite{ND}. In particular, variation of $S$ with respect to the inverse metric generates (\ref{MEQ}) and the constraint
\begin{equation}\label{C}
	\nabla_{\alpha}(u^{\alpha}u^{\beta})=0.
\end{equation}
\par From the definition of $\varPhi_{\alpha\beta}$ given by (\ref{Phiab}), the first term of $\pounds_{X}(u_{\alpha}u_{\beta})$, $X^{\lambda}\nabla_{\lambda}(u_{\alpha}u_{\beta})$, vanishes in an affine parameterization since the the collinearity $X^{\lambda}=fu^{\lambda}$ generates the geodesic terms $u^{\lambda}\nabla_{\lambda}u_{\alpha}$ and $u^{\lambda}\nabla_{\lambda}u_{\beta}$ that both vanish. Moreover, the variation generates a term $-X^{\lambda}\nabla_{\lambda}(u_{\alpha}u_{\beta}) $ as shown in Appendix B of Ref. \cite{ND}, which cancels the leading term in $\pounds_{X}(u_{\alpha}u_{\beta}) $. Thus, the Lagrangian formulation eliminates the geodesic terms in $\varPhi_{\alpha\beta}$ and it can be expressed as
\begin{equation}
	\varPhi_{\alpha\beta}=\frac{1}{2}(\nabla_{\alpha}X_{\beta}+\nabla_{\beta}X_{\alpha})+u^{\lambda}(u_{\alpha}\nabla_{\beta}X_{\lambda}+u_{\beta}\nabla_{\alpha}X_{\lambda}).
\end{equation}
\par It is important to note that the constraint (\ref{C}) can be obtained from the proof of the ODT before doing any variations of the action functional, and $\Phi$ has the global property 
\begin{equation}\label{intPhi}
	\int \Phi\sqrt{-g}d^{4}x=\int\nabla_{\alpha}X_{\beta}(g^{\alpha\beta}+2u^{\alpha}u^{\beta})\sqrt{-g}d^{4}x=0.
\end{equation}Despite that this integral vanishes, it has the metric variation $\delta\int\Phi\sqrt{-g}d^{4}x=-\int\varPhi_{\alpha\beta}\delta g^{\alpha\beta}\sqrt{-g}d^{4}x, $ which cannot be obtained by dismissing the first term $\nabla_{\alpha}X^{\alpha}$ as a total divergence and then performing the variation on the second term.
\par   The action $
S^{EHG}=\frac{c^{3}}{16\pi G}\int (R-\Phi)\sqrt{-g}d^{4}x $ that is obtained by adding (\ref{intPhi}) to the Einstein-Hilbert action generates the modified Einstein equation with no cosmological constant. If $ \Phi$ is set to the constant 2$\Lambda $, the Einstein equation with a cosmological constant is obtained accordingly, which contradicts (\ref{intPhi}). Thus, $ \Phi $ dynamically replaces the cosmological constant and the \emph{complete} Einstein equation \begin{equation}\label{ME}
	G_{\alpha\beta}+\varPhi_{\alpha\beta}=\frac{8\pi G}{c^{4}}\tilde{T}_{\alpha\beta}
\end{equation} is equivalent to (\ref{EIN}) with
\begin{equation}\label{T}
	T_{\alpha\beta}=\tilde{T}_{\alpha\beta}-\frac{c^{4}}{8\pi G}\varPhi_{\alpha\beta}. 
\end{equation}
The local conservation law, $\nabla^{\alpha}T_{\alpha\beta}$ = 0, follows from the diffeomorphic invariance of MGR as discussed in Ref. \cite{N}; it is consistent with  Einstein's fundamental postulate. In the vacuum, it follows from (\ref{T}) that $ \nabla^{\alpha}\varPhi_{\alpha\beta}=0$ and the contracted Bianchi identity is satisfied from $G_{\alpha\beta}+\varPhi_{\alpha\beta}=0 $ as required. \par Variation of the action functional $S$ with respect to $u^{\nu}$ generates
\begin{equation}\label{u}
	u_{\nu}=\frac{\partial_{\nu}f}{\Phi}
\end{equation}and variation with respect to $f$, the magnitude of $X^{\alpha}$, yields the constraint
\begin{equation}\label{nu}
	\nabla_{\alpha}u^{\alpha}=0,
\end{equation}which is equivalent to $\pounds_{u}\Phi=0$ using (\ref{u}) and $\square f=0$ from (\ref{con}) in Appendix A.
\section*{Some ramifications of MGR}
From this review of MGR, it follows that exposing the line element vector field in $\varPhi_{\alpha\beta}$ leads to new solutions for the metric from (\ref{ME}) that depend on the line element covectors in the Lorentzian metric (\ref{gab}). The line element field is a line subbundle $l$ of the tangent bundle on the Lorentzian manifold $\mathcal{M}$. The myriad of covectors that could belong to $ l$ are restricted to those of (\ref{u}) constrained by (\ref{nu}).  The line subbundle gives MGR the extra freedom to describe dark matter as presented in Ref. \cite{ND}. The extended Schwarzschild solution to the Einstein equation in the vacuum of a Lorentzian spacetime was obtained in terms of the line element covectors of a particular Lorentzian metric. The power four Tully-Fisher relation was derived; gravitational lensing, gravitational time delay, and other cosmological implications of dark matter were explained in terms of MGR without adding any dark matter profile. For example, the calculation of the total advance of the perihelion of Mercury due to GR, dark matter from MGR, and the quadrupole moment of the Sun yielded 43.0121 arcsec/century, which compared excellently to the measured result in Ref. \cite{Pir} of $43.0115 \pm 0.0085$ arcsec/century. \par Whereas Einstein developed GR in the tangent bundle on a paracompact Riemannian manifold where a Riemannian metric and a timelike regular vector exist. GR describes the gravitation of ordinary matter; it does not depend explicitly on \textbf{X}.  
\par Since $\varPhi_{\alpha\beta}$ is local and measurable, one would expect to verify its presence in every locally flat reference frame. That follows from the unique connection-independent property of $\varPhi_{\alpha\beta}$ from its construction in terms of the Lie derivative; it does not depend on the Levi-Civita connection so its value in flat or curved spacetimes is the same. Although the specific terms in the Lie derivative of a tensor are coordinate-dependent, the sum of all terms in the expression is coordinate-independent. The Levi-Civita connection in the Lie derivative essentially plays the role of choosing a particular coordinate system with partial derivatives replacing the covariant derivatives. The local gravitational energy $\varPhi_{00}$ is now calculated in two different coordinate systems from (\ref{Phi00}).\par In a maximally symmetric coordinate system described by the Friedmann-Lema\^{i}tre-Robertson-Walker (FLRW) metric with $\kappa=1$, the gravitational energy from Ref. \cite{ND} is
\begin{equation}
	\varPhi_{00}=\Lambda+\frac{8\pi G\varrho}{c^{2}}-\frac{3}{a^{2}}-\frac{6}{c^{2}}\int \frac{\dot{H}}{a}da	
\end{equation}where $\Lambda$ is the cosmological constant of measured value $\Lambda=1.1\times10^{-52}m^{-2}$, $\varrho$ is the ordinary mass density presently observed to be $\simeq3\times10^{-27}kg\;m^{-3}$ and $a$ is the cosmological scale factor that has the value $4.36\times10^{26}m$ today. The measureable gravitational energy is very small being $\sim10^{-52}m^{-2}$. In a comoving frame with $u^{0}=1,\;u_{0}=-1$,  $ (1+2u_{0}u^{0})=-1$ and the gravitational energy density $E=-\frac{c^{4}}{8\pi G}\varPhi_{00}$  is 
\begin{equation}\label{E}
	E=\frac{c^{4}}{8\pi G}\Lambda+c^{2}\varrho-\frac{3c^{4}}{8\pi Ga^{2}}-\frac{3c^{2}}{4\pi G} \int \frac{\dot{H}}{a}
\end{equation} The first term is $\varrho_{\Lambda}=5.32\times10^{-10}\frac{J}{m^{3}}$ and the second term today is the ordinary matter energy density of $ 2.7\times10^{-10} \frac{J}{m^{3}}$. The third term represents the vacuum and has the present value $\varrho_{vac}=7.62\times10^{-11} \frac{J}{m^{3}}$. The fourth term represents the flow of gravitational energy density through the cosmos in terms of $\dot{H}$ and has the approximate value today of $-1.05\times10^{-8}\frac{J}{m^{3}}$.\par In the static spherical coordinate system of Ref.\cite{ND}, $\partial_{0}X_{\alpha}=0$ and $E=-(1+2u_{0}u^{0})(\frac{GM^{2}}{4\pi r^{4}}-\frac{c^{4}a_0^{2}}{16\pi G })$. The static condition $\partial_{0}X_{\alpha}=0$ is satisfied by $u_{\alpha}\partial_{0}f=-f\partial_{0}u_{\alpha} $ from the collinearity $X_{\alpha}=fu_{\alpha}$; $u_{0}u^{0}$ does not vanish and $u_{\alpha}u^{\alpha}=-1$ holds with $-1\leq u_{0}u^{0}<0$. If this static spherically symmetric system satisfies $u_{0}u^{0}=u_{i}u^{i}\;\;i=1,2,3 $\:\:no sum on i, then $u_{0}u^{0}=-\frac{1}{4}$. The radial term equals the Newtonian gravitational energy density and that calculated in GR from the weak field approximation: $-\frac{GM^{2}}{8\pi r^{4}}\frac{J}{m^{3}} $, which has the value -$2.25\times10^{4}\frac{J}{m^{3}}$ on the Earth relative to the Sun. Using the value of $a_{0}=5.737\times10^{-41}$ for the calculation of the dark matter of the Sun in Ref.\cite{ND}, the second term in brackets has the value $-7.96\times10^{-39}\frac{J}{m^{3}}$, which is
generally too small to be measurable. \par In a comoving frame, $ (1+2u_{0}u^{0})=-1$, and $E$ is positive and twice both the Newtonian and GR values. This is not surprising because in GR, it is calculated from the Newtonian potential $\varphi$ in terms of a minute change in the metric relative to the flat spacetime Minkowski metric in the linearized field equations. The infinite sum of higher than first order perturbations in the calculation of the gravitational energy density apparently contribute the same amount as the first order contribution to E. However, there is no way of knowing that in GR because it has no tensor that explicitly represents the energy-momentum of the gravitational field. That $E$ is positive in a comoving frame is necessary to describe the local energy density of a gravitational wave: $E=(\frac{GM^{2}}{4\pi r^{4}}-\frac{c^{4}a_0^{2}}{16\pi G })\frac{J}{m^{3}}$.  
\par There are many modifications of GR obtained by adding a tensor $h_{\alpha\beta}$ to the ordinary matter tensor so that the Einstein equation holds: 
\begin{equation}\label{GMR}
	G_{\alpha\beta}+\Lambda g_{\alpha\beta}=k(\tilde{T}_{\alpha\beta}+h_{\alpha\beta}) 	
\end{equation}
where $h_{\alpha\beta}$ consists of a linear sum of divergenceless non-Lovelock tensors and the Lovelock tensors in a spacetime of greater than four dimensions. For example, the non-trivial Codazzi tensors $\xi_{\beta\lambda} $ satisfy  $\nabla_{\alpha}\xi_{\beta\lambda}=\nabla_{\beta}\xi_{\alpha\lambda}$. The tensor $H_{\alpha\beta}=\xi_{\alpha\beta}-g_{\alpha\beta}\xi_{\mu}^{\mu}$ with $\nabla^{\alpha}\xi_{\alpha\beta}=g_{\alpha\beta}\nabla^{\alpha}\xi_{\mu}^{\mu} $ can be constructed \cite{Mant} from the Codazzi tensors. In GR, $\tilde{T}_{\alpha\beta}$ is divergenceless and that property of $H_{\alpha\beta} $ ensures $\nabla^{\alpha}(\tilde{T}_{\alpha\beta}+H_{\alpha\beta})=0.$ Many theories of gravity can be described from a suitable choice of the Codazzi tensors \cite{MantF} in the FLRW metric, including $f(R)$, Gauss-Bonnet $f(G)$, teleparallel $f(T)$, Lovelock gravity, Einsteinian cubic $f(P)$, and Conformal Killing gravity (CKT). Tensor-vector-scalar modifications of GR have the same form as (\ref{GMR}) but with additional equations of motion for the added tensor, vector and scalar.\par However, $\varPhi_{\alpha\beta}$ is missing in (\ref{GMR}); the complete Einstein equation is stated by (\ref{ME}). The ODT guarantees that $\varPhi_{\alpha\beta}$ is the \emph{only} tensor that is orthogonal to \emph{all} divergenceless tensors in $h_{\alpha\beta}$  when $\tilde{T}_{\alpha\beta}\neq0$. No divergenceless non-Lovelock tensor or higher dimensional Lovelock tensor is connection-independent. That property of $\varPhi_{\alpha\beta}$ is unique relative to all other modifications of GR, which sets MGR apart from competing theories of gravity.
\par Having reviewed the line element vector field and its fundamental importance to $\varPhi_{\alpha\beta}$ of MGR, this paper now proceeds as follows: In Section 2, the Lie derivative of the metric is introduced into quantum theory from the symmetrization of the spin-1 KG equation. It shows how the line element field of a particular Lorentzian metric, the Lie derivative of that metric along the flow of the line element vector that belongs to that metric, and the Ricci tensor constructed from that metric describes the spin-1 Proca equation in curved spacetime, which unifies gravity and the spin-1 field. Section 3 investigates the conservation of charge and provides the derivation of the spin-0 KG wave equation for the scalar $\Phi$ from the line element vector field. In Section 4, properties of $\Phi$ and the Lorenz constraint are presented, and the global conformal scalar curvature of spacetime is developed from the geometrical description of the Proca equation. Section 5 is a discussion of singularities in a spherically symmetric Lorentzian spacetime. Section 6 explores quantum entanglement, the wave-particle duality, and the existence of gravitons from a geometric point of view. Section 7 presents the unification of gravity with quantum field theory (QFT) by embedding a four-dimensional background-independent Lorentzian manifold into a higher-dimensional Minkowskian manifold where QFT exists.
\section{Unifying the spin-1 Proca equation with gravity}  Unifying spin-1 bosons with gravity requires the implementation of a regular vector field that satisfies the spin-1 Proca equation in curved spacetime and is a fundamental part of a particular Lorentzian metric. That vector field $X^{\beta}$ is taken to be one of the pair of vectors in the line element field, which exists in all Lorentzian metrics. \footnote{It should be noted that since the Proca equation is linear in $X^{\beta}$, a constant times $ X^{\beta}$ defines a new vector with different units. This allows the line element vector to be used in the Proca equation with $k=0$ as the electromagnetic vector potential.} 
\par It is now shown how the line element field vector is introduced into quantum theory from the Lie derivative of the Lorentzian metric along the line element vector that belongs to a particular Lorentzian metric. First, in flat Minkowski spacetime, spin-1 neutral bosons described by the vector field $X^{\beta}$ (not considered to be a line element vector at this point of the discussion) obey the Proca equation $\partial_{\alpha}K^{\alpha\beta}=k^{2}X^{\beta}$  with the Lorenz constraint $\partial_{\alpha}X^{\alpha}=0$, where $ k=\frac{m_{0}c}{\bar{h}} $, $ m_{0} $ is the rest mass attributed to a particular spin-1 particle, and $K^{\alpha\beta}=\partial^{\alpha}X^{\beta}-\partial^{\beta}X^{\alpha} $. That is equivalent to the spin-1 multi-spin wave equation \cite{Tak} $\partial^{\alpha}\partial_{\alpha}X^{\beta}=k^{2}X^{\beta}$ with the Lorenz constraint. In curved spacetime, the corresponding asymmetric wave equation for the spin-1 vector field $X^{\beta}$ minimally coupled to gravity is 
\begin{equation}\label{asymKG}
	\square X^{\beta}=k^{2}X^{\beta}
\end{equation} where $\square:=\nabla_{\alpha}\nabla^{\alpha}$. However, the Proca equation $ \nabla_{\alpha}K^{\alpha\beta}=k^{2}X^{\beta} $ with the Lorenz constraint $ \nabla_{\alpha}X^{\alpha}=0 $ is not a wave equation because covariant derivatives do not commute: $\nabla_{\alpha}K^{\alpha\beta}=\square X^{\beta}-\nabla_{\alpha}\nabla^{\beta}X^{\alpha}$. \par The symmetrization of (\ref{asymKG}) with the covector $X_{\beta}$, as first discussed in Ref. \cite{NQ}, introduces  the Lie derivative of the metric along the flow of a line element vector into quantum theory:
\begin{equation}\label{Sym}
	\square X_{\beta}=\frac{1}{2}\nabla^{\alpha}(\tilde{\Psi}_{\alpha\beta}+K_{\alpha\beta})
\end{equation}where $\tilde{\Psi}_{\alpha\beta}=\nabla_{\alpha}X_{\beta}+\nabla_{\beta}X_{\alpha}=\pounds_{X}g_{\alpha\beta}  $ and $ K_{\alpha\beta}=\nabla_{\alpha}X_{\beta}-\nabla_{\beta}X_{\alpha} $. The Lorenz constraint is replaced by $-\Phi$, which follows from the collinearity $X^{\alpha}=fu^{\alpha}$, $ u^{\beta}\nabla_{\alpha}u_{\beta}=0$, (\ref{u}), and (\ref{nu}):
\begin{equation}\label{Phi}
	\Phi=\nabla_{\alpha}X^{\alpha}+2u^{\alpha}u^{\beta}\nabla_{\alpha}X_{\beta}=-\nabla_{\alpha}X^{\alpha}\neq0.
\end{equation}
The symmetrization invokes the Lie derivative of the metric, $\pounds_{X}g_{\alpha\beta}$, and incorporates the physics of the Proca equation. The Lie derivative is taken along the line element vector field that belongs to a particular Lorentzian metric.
\par The Proca equation with a current $J_{\beta}$ in curved spacetime is
\begin{equation}\label{P}
	\nabla^{\alpha}K_{\alpha\beta}=k^{2}X_{\beta}+J_{\beta}.
\end{equation}Its left-hand side can be expressed as 
\begin{equation}\label{P1}
	\begin{split}
		\nabla^{\alpha}K_{\alpha\beta}=\square X_{\beta}-R^{\lambda}_{\beta}X_{\lambda}+\nabla_{\beta}\Phi
	\end{split}
\end{equation}using the commutation relation $[\nabla_{\beta},\nabla_{\mu}]X_{\alpha}=-R^{\lambda}\,_{\alpha\beta\mu}X_{\lambda}$ and (\ref{Phi}). The symmetrization (\ref{Sym}) provides additional structure from which to define the current 
\begin{equation}\label{J}
	J_{\beta}:=\nabla_{\beta}\Phi-R^{\lambda}_{\beta}X_{\lambda} 	
\end{equation}geometrically. Then it follows from (\ref{Sym}) that
\begin{equation}\label{DPsi1}
	\nabla^{\alpha}\tilde{\varPsi}_{\alpha\beta}=\square X_{\beta}-J_{\beta}
\end{equation}and (\ref{asymKG}) holds. The symmetrization preserves the fundamental wave nature of the spin-1 field within the structure of the Proca equation. 
\par From (\ref{Sym}), it follows that the Lie derivative of a particular Lorentzian metric (\ref{gab}) along the flow of a line element vector $X^{\beta}$ whose covector belongs to that metric, demands the Proca tensor $K_{\alpha\beta}$ and the gravitational energy-momentum tensor $\varPhi_{\alpha\beta}$, which contains the Lie derivative of that particular Lorentzian metric, to be constructed from the same vector. The Lie derivative of the chosen Lorentzian metric is always along the line element vector associated with that metric.\par 
The line element vector that is rudimentary to a Lorentzian metric in the geometrical description of gravity is a fundamental quantum vector because it satisfies both the spin-1 KG wave equation and the Proca equation in curved spacetime. The Proca equation and the complete Einstein equation inherit the timelike nonvanishing properties of the line element field. The Lie derivative of the Lorentzian metric is the entity that links gravity with quantum theory. This starkly contrasts traditional quantum field theory that has not embraced the symmetric part of the spin-1 wave equation, leaving the Proca equation unrelated to a pure wave equation in curved spacetime, and gravity independent of quantum theory.
\par A geometrical description of the Proca equation in terms of the Lie derivative of the metric follows from (\ref{P1}) and (\ref{asymKG}):
\begin{equation}\label{Pg}
	\nabla_{\alpha}K^{\alpha\beta}=k^{2}X^{\beta}+\nabla^{\beta}\Phi-R_{\lambda}^{\beta}X^{\lambda}
\end{equation}where
\begin{equation}\label{PhiL}
	\Phi=-\frac{1}{2}g^{\alpha\beta}\pounds_{X}g_{\alpha\beta}.
\end{equation}
The line element vector field and the collinear unit covectors of the Lorentzian metric, the Lie derivative of the metric along the flow of one of the pair of vectors in the line element field, and the Ricci tensor constructed from the metric geometrically describe the spin-1 Proca equation in curved spacetime. The curvature of spacetime due to gravity is determined by the Einstein tensor, which is constructed from the Lorentzian metric and its first two derivatives. Thus, the Lorentzian metric and its inherent line element field geometrize gravity and the spin-1 Proca equation in curved spacetime. Moreover, the Lie derivative of a particular Lorentzian metric along the flow of the line element vector links the Proca equation to the Einstein equation in curved spacetime, which unifies gravity and the spin-1 field. \par This unification holds in both the micro and macro worlds. It is well known that electrodynamics with Maxwell's equations has been successfully tested on cosmic scales and in the micro world of quantum theory. Similarly, the line element vector field associated with a particular Lorentzian metric has the same properties in both the micro and macro worlds; the spherically symmetric Lorentzian microworld in MGR is continuous and singularity-free to the Planck length as proved in Section 5, and the same metric applied to the cosmos generated excellent results in Ref. \cite{ND} where dark matter is described from $\varPhi_{ii}$ without a dark matter profile.\par The action functional $S^{U}$ for the unification of gravity and the spin-1 field is presented in Appendix A from which the complete Einstein equation and the spin-1 wave equation are shown to satisfy the variational equation obtained from the variation of $S^{U}$ with respect to a line element unit vector $u^{\nu}$.
\par Since a spin-1 vector is equivalent to an outer product involving a spin-1/2 Dirac spinor $\varPsi$ and its Hermitian conjugate $\varPsi^{\dagger}$ by the relation $X^{\beta}=\varPsi^{\dagger}\gamma^{0}\gamma^{\beta}\varPsi $ where $\gamma^{\beta}$ are the Dirac gamma matrices, the KG wave equation (\ref{asymKG}) and thus (\ref{Pg}) and its geometrical properties hold for a spin-1 vector boson and a pair of spin-1/2 fermions.

\section{Conservation of charge and the spin-0 KG wave equation} 
Setting $k=0$ in (\ref{P}) gives the Maxwell equation in curved spacetime
\begin{equation}\label{M1}
	\nabla^{\alpha}K_{\alpha\beta}=J_{\beta}
\end{equation}and from the definition of $K_{\alpha\beta}=\nabla_{\alpha}X_{\beta}-\nabla_{\beta}X_{\alpha}$, the second Maxwell equation is
\begin{equation}\label{M2}
	\nabla_{\mu}K_{\alpha\beta}+\nabla_{\alpha}K_{\beta\mu}+\nabla_{\beta}K_{\mu\alpha}=0.
\end{equation}Charge is conserved, which follows from the commutator of the covariant derivative acting on $K^{\alpha\beta}$:\newline ($\nabla_{\alpha}\nabla_{\beta}-\nabla_{\beta}\nabla_{\alpha})K^{\alpha\beta}=-2R_{\alpha\beta}K^{\alpha\beta}=0$ since $K^{\alpha\beta}$ and $R_{\alpha\beta}$ have opposite symmetries. That requires $\nabla_{\beta}\nabla_{\alpha}K^{\alpha\beta}=0$, which demands $\nabla_{\beta}J^{\beta}=0$. \par If $k\neq0$, $\nabla_{\beta}\nabla_{\alpha}K^{\alpha\beta}=0$ requires
\begin{equation}\label{J1}
	\nabla_{\beta}J^{\beta}=k^{2}\Phi
\end{equation} and charge is not locally conserved. However, it is globally conserved, which follows from the unique global property of $\Phi$ from equation (\ref{intPhi}): $k^{2}\int\Phi\sqrt{-g}d^{4}x=0.$ \par  The commutation relation of the line element quantum vector
\begin{equation}\label{CRS0}
	[\square,\nabla_{\beta}]X^{\beta}=\nabla^{\alpha}[\nabla_{\alpha},\nabla_{\beta}]X^{\beta}+[\nabla^{\alpha},\nabla_{\beta}]\nabla_{\alpha}X^{\beta}
\end{equation}leads to the spin-0 KG equation for the scalar $\Phi$. The second term in (\ref{CRS0}) vanishes because $[\nabla_{\lambda},\nabla_{\beta}]\nabla^{\lambda}X^{\beta}=R_{\sigma\beta}(\nabla^{\sigma}X^{\beta}-\nabla^{\beta}X^{\sigma})=0$, and the first term is equivalent to $-\nabla^{\alpha}(R_{\lambda\alpha}X^{\lambda})$, which means $\int[\square,\nabla_{\beta}]X^{\beta}\sqrt{-g}d^{4}x=\int(-\square \Phi+k^{2}\Phi)\sqrt{-g}d^{4}x=0$ from (\ref{asymKG}) and (\ref{Phi}). Thus, the spin-0 wave equation
\begin{equation}\label{KG0}
	\square \Phi=k^{2}\Phi
\end{equation}is a solution with $\Phi\neq0$ since $X^{\beta}$ is not a Killing vector.  \par It should be noted that the spin-0 wave equation with $\Phi$ defined by (\ref{PhiL})
is consistent with the definition of $J^{\beta}$ and (\ref{J1}):
\begin{equation}\label{JP}
	\begin{split}
		0=\int\nabla_{\beta}(\nabla^{\beta}\Phi-R^{\lambda\beta}X_{\lambda})\sqrt{-g}d^{4}=\int\square\Phi\sqrt{-g}d^{4}
	\end{split}
\end{equation}after integrating by parts, so $\int(\square\Phi-k^{2}\Phi)\sqrt{-g}d^{4}=0. $ 
\par The line element vector belonging to a particuar Lorentzian metric that describes the Proca tensor also produces the spin-0 wave equation from the relation (\ref{Phi}). Thus, any discussion about unifying the spin-1 Proca equation with gravity automatically includes spin-0 particles. For clarity, although the inverse Compton wavelength $k$ is generally not the same for particles of different spins, the same symbol $k$ is used for different spins where there should be no confusion. 

\section{Properties of $\Phi$ and the Lorenz constraint}
Properties of $\Phi$ are now investigated. One  question immediately arises: is the scalar $\Phi$ large enough to measurably affect local charge conservation? In the vacuum, (\ref{ME}) generates $R=\Phi\neq0$, but in Minkowski spacetime, ten Killing vectors of the Poincar\'{e} group exist that render $\Phi=0$ everywhere; there is no gravitational influence in a flat Minkowski spacetime. However, the existence of gravitational energy-momentum at every point in spacetime produces a gravitational field from $\varPhi_{\alpha\beta}$. Thus, spacetime is not Minkowskian and although $\Phi$ does not vanish, it can be minute. For example, in a region of spacetime that is free of ordinary matter and described by the spherical metric in Ref. \cite{ND}, $\Phi=\frac{6a_{0}}{r}-\frac{2b(3+2\ln r)}{r^{2}}  $ where $b=\pm\sqrt{\frac{\mid a_{0}\mid GM}{c^{2}}}$, $M$ is the gravitating mass composed of both ordinary and dark matter, and $a_{0}>0$ is a solution to $v^{2}=\frac{GM}{r}+c\sqrt{a_{0}GM}-\frac{a_{0}c^{2}r}{2}$ for the orbital velocity $v$ of a star at a radial distance $r $ from the center of mass of a spherical galaxy. If we conceptually apply this to the Sun-Earth system and ignore the eccentricity of the orbit of the Earth around the Sun, $a_{0}\simeq2.81\times10^{-19}m^{-1}$, $b=-2.04\times10^{-8}$, and  $\Phi=1.14\times10^{-28}\,m^{-2} $. Thus, the Proca equation with the Lorenz constraint $\partial_{\alpha}X^{\alpha}=0$ in a flat background spacetime on the Earth is an excellent approximation to the Proca equation in the presence of gravity. Nevertheless, it must be emphasized that $\Phi\neq0$. 

\par  From the geometrical description of the Proca equation, it follows that $\Phi^{2}$ is responsible for the global conformal scalar curvature of spacetime from charged and neutral spin-1 particles. A Lorentzian metric (\ref{gab}) demands a product of line element covectors to be expressed as $X_{\alpha}X_{\beta}=\frac{f^{2}}{2}(g^{+}_{\alpha\beta}-g_{\alpha\beta}) $, which represents the conformal metric $\bar{g}_{\alpha\beta}=\frac{f^{2}}{2}(g^{+}_{\alpha\beta}-g_{\alpha\beta})$. The conformal Ricci scalar $\bar{R}$ is $\bar{R}=R^{\lambda\beta}X_{\lambda}X_{\beta}$. If $J^{\beta}=0$, $ \bar{R}=X_{\beta}\nabla^{\beta}\Phi$ and it follows from (\ref{Phi}) after integrating by parts that 
\begin{equation}
	\int \bar{R}\sqrt{-g}d^{4}x=\int \Phi^{2}\sqrt{-g}d^{4}x>0.
\end{equation}Neutral spin-1 bosons and pairs of neutrinos demand a positive global conformal scalar curvature. If $J^{\beta}<0$,  $\int \bar{R}\sqrt{-g}d^{4}x>\int\Phi^{2}\sqrt{-g}d^{4}x>0$; if $J^{\beta}>0$, $\int \bar{R}\sqrt{-g}d^{4}x<\int\Phi^{2}\sqrt{-g}d^{4}x$. These results are independent of $k$ so $\Phi^{2}$ is responsible for the global conformal scalar curvature from all massive or massless particles, including dark matter.
\section{Singularities in a spherically symmetric Lorentzian spacetime}
The Kretschmann invariant $R_{\alpha\beta\gamma\rho} R^{\alpha\beta\gamma\rho}$ in the Schwarzschild metric of GR is $\mathcal{K}=\frac{48G^{2}M^{2}}{c^{4}r^{6}}$, which proves there is a true spacetime singularity at $r=0$. GR depends explicitly on the Riemannian part of the Lorentzian metric and not on the line element vectors so $r$ vanishes in GR. Although it is generally speculated that quantum gravitational effects are required \cite{Alesci} to smooth out the classical spacetime singularity in black holes predicted by GR, that is not the case in MGR because solutions to the Einstein equation depend explicitly on both of the Riemannian metric and the line element vectors that are introduced into the Einstein equation from $\varPhi_{\alpha\beta}$. \par The regular (everywhere nonvanishing) covectors $X_{\alpha}=\frac{f\partial_{\alpha}f}{\Phi} $ from (\ref{u}) define the time oriented line subbundle in MGR. $\Phi$ is nonvanishing iff $\varPhi_{\alpha\beta}$ does not contain a Killing vector, which is always the case unless a particular symmetry is involved. 
\par In the spherically symmetric spacetime free of ordinary matter in MGR \cite{ND}, the extended Schwarzschild solution to the Einstein equation is $e^{-\lambda}=-a_0r+2b\ln r+\frac{c_{1}}{r}+c_{2}$ where $c_{1}=a_{2}$, $a_{1}=-2b$, $c_{1}=-\frac{2GM}{c^{2}}$, and $c_{2}=1$. Choosing $f=\Phi$ and  
\begin{equation}\label{f}
	f=-e^{-\lambda}
\end{equation}
requires $X_{1}=\partial_{1}f=a_{0}+\frac{a_{1}}{r}+\frac{a_{2}}{r^{2}}$, which is the radial covector for the extended Schwarzschild solution. The magnitude $f$ of the line element covector $X_{1}$ can not vanish and $f$ must be finite for solutions to the Einstein equation to be physically acceptable, which is the case since \ref{intPhi} demands $\Phi$ to be bounded. Moreover, all components of $u_{\alpha}u^{\alpha}=-1$ such as $g^{11}(u_{1})^{2}=-f(\partial_{1}\ln f)^{2}$ are bounded so $f$ is bounded. Thus, $0<r<\infty$.
\par 
The Kretschmann invariant for the extended Schwarzschild solution of MGR is
\begin{equation}\label{key}
	\begin{split}
		\mathcal{K}=\frac{48G^{2}M^{2}}{c^{4}r^{6}}+\frac{16bGM(3-2\ln r)}{c^{2}r^{5}}+\frac{4b^{2}(5+4\ln^{2}r)}{r^{4}}\\-\frac{16a_{0}b(1+\ln r)}{r^{3}}+\frac{8a^{2}_{0}}{r^{2}},
	\end{split}
\end{equation}
which does not blow up because $r>0$, and the logarithmic terms $\frac{\ln r}{r^{5}}$, $\frac{\ln^{2}r}{r^{4}}$, and $\frac{\ln r}{r^{3}}$ are extremely small when $r$ is enormously large but finite. There is no need for a quantum gravitational effect or any other entity to account for a perceived singularity at $r=0$ in MGR because it is free of true singularities in the extended Schwarzschild metric. Spacetime is continuous and singularity-free to the Planck length.
\section{Quantum theory and the unified spin-1 field}
Establishing the unification of gravity and the spin-1 field leads to the implication of MGR to other aspects of quantum theory. In that regard, there are three fundamental quantum attributes, first discussed in Ref. \cite{NQ} and now expanded, to be reconciled relative to the foundations of MGR in terms of the line element field and the Lie derivative of the metric, which have been overlooked in quantum theory; they can be described geometrically and are not unique to quantum theory. 
\subsection{Quantum entanglement}
Einstein's famous description of quantum entanglement \cite{EPR} in 1935 as ``spooky action at a distance" is typically understood today as correlation without communication between separated particles where those particles are considered as one object. Many experiments \cite{Friis} have conclusively determined that quantum entanglement is a global property presumed to be unique to quantum theory. However, quantum theory has neglected the Lie derivative of the metric in the spin-1 wave equation, and there may be superluminal communication between entangled particles after all. Measurements of the ``speed of spooky action" have been made \cite{Juan,Amato} with the most recent bound of about $3.3\times10^{4}c$. The speed of the transfer of quantum information in quantum entanglement can be described geometrically, which follows from the structure of the spin-1 wave equation and the group of diffeomorphisms Diff($\mathcal{M}$).  \par The line element covector $X_{\beta}$ is a solution \cite{fland} to the spin-1 wave equation $\square X_{\beta}=k^{2}X_{\beta}$ from which the anti-symmetric Proca tensor $K_{\alpha\beta}$ and its symmetric partner $\pounds_{X}g_{\alpha\beta}$ are constructed. The Proca tensor describes the quantum properties of the particle nature of the spin-1 field. The Lie derivative of the metric contains all quantum information about the spin-1 particle from the solution $X_{\beta}$ to the wave equation.

\par The Lagrangians for the field equations of particles with spins 0,1,1/2 are scalars. Any action functional with a scalar Lagrangian constructed from the tensor fields of the Lagrangian is invariant under the group Diff($\mathcal{M}$) of diffeomorphisms. Since the field equations of spin-0,1 bosons and a pair of spin-1/2 fermions involve the Lie derivative of the metric, they have a geometrical interpretation in terms of a family of diffeomorphisms. Given a diffeomorphism $ \phi: \mathcal{M}\longrightarrow \mathcal{M} $, the Lie derivative of the metric is constructed from the pullback $\phi_{t\ast}$ of the metric under Diff($\mathcal{M}$):\newline $
\pounds_{X} g_{\alpha\beta}= \lim_{t \to 0} \{\frac{\phi_{t\star}[g_{\alpha\beta}(\phi_{t}(p))]-g_{\alpha\beta}(p)}{t}\}$
where $\phi_{t}  $ is the flow down the integral curves defined as those curves $x^{\beta}(t)$ which solve $ X^{\beta}=\dfrac{dx^{\beta}}{dt} $ for the line element vector field $X^{\beta}$ where $ t $ defines a one-parameter family of diffeomorphisms. The Lie derivative of the metric tells us how fast the metric changes as it moves along the integral curves. Moreover, the Lorentz group is a subgroup of Diff($\mathcal{M}$) so the pullback of the metric is not restricted to the Lorentz group. The metric at a point $p^{\prime}$ on the integral curve far from a given point $p$ on that curve can be pulled back
superluminally to the neighborhood of $p$, or the inverse metric can be pushed forward from $p$ to $p^{\prime}$ with $(\phi_{t}^{-1})^{\star}$. Although there appears to be an upper limit of how fast a Lorentzian metric can travel along an integral curve, it is possible for quantum information to be transmitted superluminally between entangled particles by the Lie derivative of the metric. The respective local/nonlocal characteristics of MGR and quantum theory no longer present an impenetrable barrier to unifying the theories.
\subsection{The wave-particle duality}
Quantum theory is formulated in a flat Minkowskian geometry with no effects of gravity whatsoever. Feynman \cite{Feyn} called the wave-particle duality the only mystery of quantum physics: ``... a phenomenon which is impossible, \emph{absolutely} impossible, to explain in any classical way, and which has in it the heart of quantum mechanics. In reality, it contains the \emph{only} mystery."\par However, quantum theory has neglected the Lie derivative of the Lorentzian metric as its link to gravity, and it can not escape gravity because even a small amount of ordinary matter minutely curves spacetime. The wave-particle phenomenon must be reconciled within the structure of a wave in curved spacetime. It is not that mysterious because particles of spins 0,1,1/2 in curved spacetime move as a wave at all spacetime coordinates and have inherent particle characteristics.\par   The KG wave equations in curved spacetime for spin-1 bosons and a pair of spin-1/2 fermions, and the spin-0 bosons are given by (\ref{asymKG}) and (\ref{KG0}), respectively. By modifying the Dirac equations in curved spacetime, the parent spin-1/2 KG equation can be expressed as a wave equation as shown in Ref. \cite{NQ}. That follows by introducing a complex scalar $\Omega$ into the Dirac equations as $(\gamma^{\mu}\nabla_{\mu}+\Omega\pm k)\Psi^{a}=0,$ where $\Omega^{2}=\frac{R}{4}$ and imposing additional commutation relations. Then the parent KG wave equation of the Dirac spinor $\square\Psi^{a}=k^{2}\Psi^{a}$ is satisfied. \par Thus, the KG multi-spin wave equation in curved spacetime $\square\Psi=k^{2}\Psi$ exists for the quantum field $\Psi$ with spins-0,1 and 1/2 where $k$ is the inverse Compton wavelength attributed to each particle of a given spin. The connection is the torsionless Levi-Civita connection for tensor fields, and for spinors, it contains the spinor affinities. All known bosons and fermions move as a wave everywhere in curved spacetime. Moreover, the flat spacetime d'Alembert wave operator $\partial_{\alpha}\partial^{\alpha}$ always exists in the KG equations for spins 0,1,1/2 in curved spacetime from the properties of a covariant derivative. That operator can be expressed in the momentum representation as $-\frac{p^{\mu}p_{\mu}}{\bar{h}^{2}}$ with the momentum vector $p^{\mu}$. Hence, spin 0,1 bosons and spin-1/2 fermions always move as a wave and their waves have inherent particle characteristics. That resolves the mystery of the wave-particle duality.
\par The wave or particle characteristics determined from an experiment are simply the result of a measurement of the particle-wave; the measurement does not mysteriously create a particle from a wave or a wave from a particle. The measurement is limited to the detection of all aspects of the particle-wave; depending on the experiment, it may appear to be more particle-like than wave-like or conversely, but an entity in the quantum microworld always has particle and wave characteristics. \par Observing the simultaneous particle and wave characteristics has been problematic. Nevertheless, entanglement resolves that problem; the deterministic wave-particle entanglement of two photons was achieved \cite{Rab,Man} in 2017. More importantly, a single self-entangled photon was observed to exhibit simultaneous wave and particle behaviors \cite{Qian}. That was the first experiment at the single particle level, which is required to test a quantum-mechanical entity acting as both a particle and a wave.
\subsection{Gravitons} Quantizing the gravitational field is the ambitious goal of one approach to quantum gravity, where the graviton is the spin-2 particle that mediates the gravitational force. However, forces in GR and MGR are described geometrically without the need for any mediating particle, such as the graviton, which questions its existence.\par Massless gravitons in flat Minkowskian spacetime are described in Ref. \cite{Tak} by the spin-2 Klein-Gordon equation $\partial_{\mu}\partial^{\mu}\chi_{\alpha\beta}=0$ where $\chi_{\alpha\beta}$ is a symmetric, divergenceless, and traceless (0,2) tensor field. The massless spin-2 particle minimally coupled to gravity \cite{NQ} demands that a graviton must satisfy the spin-2 KG wave equation
\begin{equation}\label{KG2}
	\square\tilde{\chi}_{\alpha\beta}=0
\end{equation} for the symmetric, divergenceless, and traceless spin-2 field $\tilde{\chi}_{\alpha\beta} $ in a four-dimensional curved spacetime. That general symmetric (0,2) tensor can be written \cite{NQ} as the sum of the ordinary matter energy-momentum tensor $\tilde{T}_{\alpha\beta}$ plus an unknown symmetric tensor $w_{\alpha\beta}$, which can be decomposed by the ODT: $\tilde{\chi}_{\alpha\beta}=a\tilde{T}_{\alpha\beta}+bw_{\alpha\beta}=a\tilde{T}_{\alpha\beta}+b(v_{\alpha\beta}+\varPhi_{\alpha\beta})$ where $a$ and $b$ are arbitrary parameters. The linear sum of symmetric divergenceless (0,2) tensors $v_{\alpha\beta}$ in the ODT is the sum of the (0,2) Lovelock tensors and those (0,2) tensors that are not Lovelock tensors, such as the trace of the Chevreton tensor that contains fourth-order derivatives and is symmetric, divergenceless, and traceless \cite{Bergq} for a source-free electromagnetic field. In a four-dimensional spacetime, the Lovelock tensors are the Einstein tensor $G_{\alpha\beta}$ and the metric $g_{\alpha\beta}$, and the non-Lovelock tensors are represented by $h_{\alpha\beta}$. Thus, $\tilde{\chi}_{\alpha\beta}=a\tilde{T}_{\alpha\beta}+b(G_{\alpha\beta}+\Lambda g_{\alpha\beta}+\varPhi_{\alpha\beta}+h_{\alpha\beta}) $. Choosing $a=-1$ and $b=\frac{c^{4}}{8\pi G}$ yields $\tilde{\chi}_{\alpha\beta}=\frac{c^{4}}{8\pi G}h_{\alpha\beta} $, which does not contain the metric as a field variable and demands $h_{\alpha\beta}$ to be traceless. The Einstein equation (\ref{MEQ}) strips the metric out of $\tilde{\chi}_{\alpha\beta}$ and is effectively hidden in a symmetric, divergenceless, and traceless (0,2) tensor field. Since the metric does not exist as a field variable in (\ref{KG2}), gravitons described by the metric do not exist. \par In a Lorentzian spacetime, the linearized Einstein equation is developed from $g_{\alpha\beta}=\eta_{\alpha\beta}+h_{\alpha\beta}$ where $h_{\alpha\beta}$ is the perturbation to the Minkowski metric. It is well-known that the Fierz-Pauli equation in flat Minkowskian spacetime that represents spin-2 particles described by the symmetric, divergenceless, and traceless non-Lovelock tensor $h_{\alpha\beta}$, is equivalent to the linearized Einstein equation in a spacetime free of ordinary matter. However, this equivalence is valid only in weak gravitational fields. Moreover, it must be emphasized that massless or massive gravitons described by the full metric do not exist in a curved Lorentzian spacetime. The ODT proves that the Einstein equation strips the full metric out of the general spin-2 field equation $\square\tilde{\chi}_{\alpha\beta}=k^{2}\tilde{\chi}_{\alpha\beta} $; the full Lorentzian metric cannot quantize spacetime. \par  As discussed in Ref.\cite{NQ}, gravitons are excluded as mediators for the force of gravity in MGR. The lack of a force mediator for gravity explains why it is a much weaker effective force than the other three known forces of nature. They have spin-1 force mediators; the massless photon for the electromagnetic force, the massive W and Z bosons for the electroweak force, and the massless gluons for the strong nuclear force. Thus, the hierarchy problem is explained within the structure of MGR.
\section{The unification of gravity and quantum field theory}
Geometrizing the spin-1 field and unifying the Proca equation with the Einstein equation by the Lie derivative of the Lorentzian metric is paramount to developing a quantum theory of gravity by the geometrization of quantum theory. That seems like a dubious statement, considering a connection to quantum field theory has not been established. QFT was created in a flat Minkowskian spacetime where there is no influence from gravity. Particles are created and annihilated in the formalism of canonical quantization, and there exists a well-defined vacuum state. Spin-2 particles in flat spacetime are described by gravitons. Nevertheless, it has been established that the full metric in curved spacetime  can not describe gravitons; gravity has no spin-2 particle representation by the full metric and cannot be quantized.  \par It is well-known \cite{Edd} that a four-dimensional pseudo-Riemannian manifold can be represented graphically as a surface of four dimensions drawn in a flat hyperspace of a sufficient number of dimensions. To unify gravity with QFT, a four-dimensional Lorentzian manifold where gravity lives with all known elementary particles must be embedded into a flat higher-dimensional manifold where QFT resides. There are many theorems on the local and global isometric embedding of a Lorentzian manifold into a higher dimensional Minkowskian manifold \cite{Fried,Greene,Clarke,MS,Min} that provide the machinery to unify gravity and QFT. The number of dimensions $N$ in the Minkowskian manifold $L^{N}$ required for the embedding varies widely; a local embedding requires $N=10$ unless some symmetry is imposed to lower the number of spatial dimensions, and a global embedding requires $N=46$ or more.  \par 
The unification of gravity with the spin-1 field and QFT has led us directly to the embedding formalism of many approaches to quantum gravity such as string theory, large extra dimensions, M theory, and other proposals as discussed in Refs. \cite{Shif,Maar,Past} (and references therein). However, these approaches to quantum gravity are background-dependent in contrast to the geometrical unification of the
spin-1 field and gravity in the four-dimensional embedded submanifold of MGR, which does not depend on the background metric. The importance of background independence in quantum gravity is of great interest as discussed in Refs. \cite{Smol,Bojo}.
\par Thus, the unification of quantum field theory and gravity is a model of quantum gravity from the geometrization of quantum theory.

\section{Conclusion}
Unifying the massive spin-1 field with gravity requires the implementation of a regular vector field that satisfies the spin-1 Proca equation and is a fundamental part of the spacetime metric. That vector field is one of the pair of vectors in the line element field (\textbf{X},-\textbf{X}), which is paramount to the existence of all Lorentzian metrics. Symmetrization of the spin-1 Klein-Gordon wave equation in curved spacetime introduces the Lie derivative of a particular Lorentzian metric along the flow of the line element vector that belongs to that metric. The Proca equation in curved spacetime can then be described geometrically in terms of the line element vector, the Lie derivative of the Lorentzian metric, and the Ricci tensor. The Lie derivative of the metric links the Proca equation with the complete Einstein equation. All known spin-0 scalar bosons, spin-1 vector bosons, and pairs of spin-1/2 fermions can be described geometrically, and are linked to the Einstein equation, which unifies those particles with gravity.
\par The structure of MGR in terms of the line element field provides important implications for quantum theory:
\begin{enumerate}
	\item An action functional with a scalar Lagrangian constructed from the tensor fields of the Lagrangian is invariant under the diffeomorphism group Diff($\mathcal{M}$). The Lie derivative of the metric along the line element vector belonging to that metric can then be expressed as a diffeomorphism, which determines the speed of the metric as it travels down the integral curve of the line element vector. The mystery of the super-luminal transfer of quantum information in quantum entanglement is explained geometrically within the structure of MGR. The respective local/nonlocal characteristics of MGR and quantum theory no longer present an impenetrable barrier to unifying the theories. 
	\item All known bosons and fermions with spins-0,1,1/2 behave as a wave with particle characteristics everywhere in curved spacetime. There is no mystery in the wave-particle duality. 
	\item The full metric belongs to the Einstein equation and cannot represent a symmetric, divergenceless, and traceless spin-2 particle that mediates the gravitational field in a four-dimensional Lorentzian spacetime; gravitons represented by the full metric as a spin-2 field variable do not exist. Spacetime cannot be quantized; it is singularity-free and continuous down to the Planck length. 
	\item Gravity has a much weaker effective force than the other three known forces of nature because it does not have force mediators determined by the full metric. This explains the hierarchy problem.
\end{enumerate}
\par The geometrization of the Proca equation and its unification with the Einstein equation by the Lie derivative of the metric provides the basis for the unification of gravity and quantum field theory. That unification is based on the embedding of a background-independent four-dimensional Lorentzian manifold containing MGR and the geometrized KG equations into a higher-dimensional Minkowskian manifold $L^{N}$ with $N\geq10$ that contains QFT. This unification is a model of quantum gravity achieved from the geometrization of quantum theory. \par Einstein believed quantum theory was not complete. He was correct because traditional quantum theory has not considered the symmetric part of the spin-1 wave equation. His dream was to construct a geometrical theory of gravity unified with quantum theory in a continuous singularity-free curved spacetime. The unification of gravity with the spin-1 field appears to fulfill that dream.
\section*{Acknowledgements}
I would like to thank the anonymous referee who provided constructive comments that improved the quality of the manuscript.
%
\appendix
\section{The action functional for the unification of gravity and the spin-1 wave equation}
The action functional for the spin-1 wave equation (\ref{asymKG}) is: $ S^{1}=-\frac{1}{2}\int[\nabla_{\alpha}X_{\beta}\nabla^{\alpha}X^{\beta}+k^{2}X_{\beta}X^{\beta}]\sqrt{-g}d^{4}x $ where $ X^{\beta} $ is an arbitrary line element vector. It is worthwhile to express $ S^{1} $ in terms of the independent variables $ f $ and $ u_{\beta} $ as:
\begin{equation}\label{S^{1}}
	\begin{split}
		S^{1}=-\frac{1}{2}\int(f^{2}\nabla_{\alpha}u_{\beta}\nabla^{\alpha}u^{\beta}-\nabla_{\alpha}f\nabla^{\alpha}f
		-f^{2}k^{2})\sqrt{-g}d^{4}x\\
		=\frac{1}{2}\int (f^{2}u_{\beta}\square u^{\beta}+\nabla_{\alpha}f\nabla^{\alpha}f
		+f^{2}k^{2})\sqrt{-g}d^{4}x
	\end{split}
\end{equation}using $ u^{\beta}\nabla_{\alpha}u_{\beta}=0 $ and integrating by parts. \par First, it is important to prove: given $ \square u^{\beta}=k^{2}u^{\beta}$, $\square X^{\beta}=k^{2}X^{\beta}$ iff $\square f=0$ and $ \nabla_{\alpha}f=\Phi u_{\alpha} $  where $X^{\beta}=fu^{\beta}$. \par The variation of $ S^{1} $ with respect to $ f $ is: 
\begin{equation}\label{dS1}
	\begin{split}
		\delta S^{1}=\int [fu_{\beta}(\square u^{\beta}-k^{2}u^{\beta})\delta f+\nabla^{\alpha}f\delta\nabla_{\alpha}f]\sqrt{-g}d^{4}x\\
		=\int [fu_{\beta}(\square u^{\beta}-k^{2}u^{\beta})-\square f)]\delta f\sqrt{-g}d^{4}x
	\end{split}
\end{equation}after integrating by parts. Since $ fu_{\beta}\neq0 $, it follows from the variation that
\begin{equation}\label{KGu}
	\square u^{\beta}=k^{2}u^{\beta}
\end{equation} provided 
\begin{equation}\label{con}
	\square f=0.
\end{equation}
\par From (\ref{u}), $ \nabla_{\alpha}f=\Phi u_{\alpha} $, so $\square(fu^{\beta})=u^{\beta}\square f+f\square u^{\beta}$ in an affine parameterization. Imposing the constraint (\ref{con}) allows the variation from (\ref{dS1}) to be expressed as
\begin{equation}\label{key}
	\int u_{\beta}[\square(fu^{\beta})-fk^{2}u^{\beta}]\sqrt{-g}d^{4}x=0
\end{equation}from which
\begin{equation}\label{KGX}
	\square X^{\beta}=k^{2}X^{\beta}
\end{equation}since $ u_{\beta}\neq0 $. $X^{\beta} $ satisfies the spin-1 KG equation provided its magnitude fulfils the homogeneous wave equation $ \square f=0 $ and (\ref{u}) holds, which from (\ref{nu}) demands $ \pounds_{u}\Phi=0 $.\par 
The unified action $S^{U}$ for the complete Einstein equation (\ref{MEQ}) of MGR and the spin-1 wave equation (\ref{asymKG}) is $S^{U}=S+S^{1}$ where $S$ is defined in (\ref{S}). Variation of $S^{U}$ with respect to $u^{\nu}$ is now performed. Unlike in GR where the line element field is ignored, the variation of the Lorentzian metric with respect to $u^{\nu}$ must be undertaken. Using $\frac{\delta S}{\delta u^{\nu}}=\frac{\delta S}{\delta g^{\alpha\beta}}\frac{\delta g^{\alpha\beta}}{\delta u^{\nu}}$, (\ref{gab}), $\delta\int\varPhi_{\alpha\beta}g^{\alpha\beta}\sqrt{-g}d^{4}x=-\int\varPhi_{\alpha\beta}\delta g^{\alpha\beta}\sqrt{-g}d^{4}x $  from \cite{ND}, and $\int f^{2}\nabla_{\alpha}u_{\beta}\nabla^{\alpha}u^{\beta}\sqrt{-g}d^{4}x=-\int f^{2}u^{\beta}\square u_{\beta}\sqrt{-g}d^{4}x$
\begin{equation}
	\begin{split}
		\frac{\delta S^{U}}{\delta u^{\nu}}=-4\int[-\frac{\tilde{T}_{\nu\beta}}{2c}+a(G_{\nu\beta}+\Lambda g_{\nu\beta}+\varPhi_{\nu\beta})]u^{\beta}\sqrt{-g}d^{4}x\\+\frac{1}{2}\int f^{2}\frac{\delta}{\delta u^{\nu}}[u^{\beta}(\square u_{\beta}-k^{2}u_{\beta})-\square f]\sqrt{-g}d^{4}x\\+\frac{1}{2}\int[f^{2}u^{\beta}(\square u_{\beta}-k^{2}u_{\beta})+\square f]g_{\alpha\beta}(u^{\alpha}\delta^{\nu}_{\beta}+u^{\beta}\delta^{\nu}_{\alpha})\sqrt{-g}d^{4}x\\=\int[-4u^{\beta}(-\frac{\tilde{T}_{\nu\beta}}{2c}+a(G_{\nu\beta}+\Lambda g_{\nu\beta}+\varPhi_{\nu\beta}))\\+\frac{f^{2}}{2}(\square u_{\nu}-k^{2}u_{\nu}+u^{\beta}\frac{\delta}{\delta u^{\nu}}(\square u_{\beta}-k^{2}u_{\beta}))\\+f^{2}u^{\beta}(\square u_{\beta}-k^{2}u_{\beta})u_{\nu}]\sqrt{-g}d^{4}x=0.
	\end{split}	
\end{equation}With $a=\frac{c^{3}}{16\pi G}$, the modified Einstein equation (\ref{MEQ}) and the spin-1 KG wave equation $\square u_{\beta}=k^{2}u_{\beta}$ satisfy this variational equation from which (\ref{KGX}) and the neutral Proca equation $\nabla^{\alpha}K_{\alpha\beta}=k^{2}X_{\beta}$  follow provided $\square f=0$.

\end{document}